\documentclass[aps,prl,twocolumn,floats,epsfig,showpacs]{revtex4-1}

\usepackage{bbm}
\usepackage{amssymb}
\usepackage{ifpdf}
\usepackage{color}
\usepackage{graphicx,epsfig}
\usepackage{amsmath}
\usepackage{amsfonts}
\usepackage{hyperref}
\usepackage{dcolumn}
\usepackage{bm}
\usepackage{xcolor}
\usepackage{graphicx}

\newcommand{\Z}{{\mathbb{Z}}}

\definecolor{red}{rgb}{0.7,0,0}
\definecolor{green}{rgb}{0.,0.35,0.}
\definecolor{blue}{rgb}{0.2,0.2,0.7} 
\definecolor{black}{rgb}{0.15,0.15,.15}

\begin{document}

\title{Atomic Quantum Simulation of $U(N)$ and $SU(N)$ Non-Abelian Lattice
Gauge Theories}
\author{D.\ Banerjee$^{1}$, M.\ B\"ogli$^{1}$, M.\ Dalmonte$^{2}$, 
E.\ Rico$^{2,3}$, P.\ Stebler$^{1}$, U.-J.\ Wiese$^{1}$, and P.\ Zoller$^{2,3}$}
\affiliation{$^{1}$Albert Einstein Center, Institute for Theoretical Physics, 
Bern University, CH-3012, Bern, Switzerland \\
$^{2}$Institute for Quantum Optics and Quantum Information of the Austrian 
Academy of Sciences, A-6020 Innsbruck, Austria \\
$^{3}$ Institute for Theoretical Physics, Innsbruck University, 
A-6020 Innsbruck, Austria}

\begin{abstract}
Using ultracold alkaline-earth atoms in optical lattices, we construct a quantum simulator for $U(N)$ and $SU(N)$ lattice gauge theories with fermionic matter based on quantum link models. These systems share qualitative features with QCD, including chiral symmetry breaking and restoration at non-zero temperature or baryon density. Unlike classical simulations, a quantum simulator does not suffer from sign problems and can address the corresponding chiral dynamics in real time.
\end{abstract}

\maketitle

\textit{Introduction.} Non-Abelian gauge fields play a central role in the dynamics of the Standard Model of particle physics.  In particular, the strong $SU(3)$ gauge interactions between quarks and gluons in Quantum Chromodynamics (QCD) give rise to the spontaneous breakdown of the chiral symmetry of the light quarks. Heavy-ion collisions produce a high-temperature quark-gluon plasma in which chiral symmetry is restored. The deep interior of neutron stars contains high-density nuclear matter or even quark matter, which may be a baryonic superfluid or a color superconductor \cite{rajagopal2000condensed}. Unfortunately, due to severe sign problems, the real-time evolution of heavy-ion collisions or the phase structure of dense QCD matter is inaccessible to first principles classical simulation methods. In condensed matter physics strongly coupled gauge theories play a prominent role in strongly correlated systems. In particular, the non-Abelian $SU(2)$ variant of quantum spin liquids has long been debated as a possible connection between the doped Mott insulator and the high-$T_c$ superconducting phase in cuprates \cite{lee2006doping}. The challenge of solving such problems motivates the development of quantum simulators for non-Abelian lattice gauge theories.  Recently, quantum simulators have been constructed for Abelian $U(1)$ gauge theories with \cite{banerjee2012,zohar2013simulating,kapit2011optical} and without coupling to matter fields \cite{zohar2012simulating,tagliacozzo2012optical}. Here, we construct a quantum simulator of $U(N)$ and $SU(N)$ strongly coupled lattice gauge theories in $(1+1)$, $(2+1)$, and $(3+1)$D using ultracold alkaline-earth (AE) atoms in an optical lattice. On the one hand, our approach is based on quantum link models (QLMs) \cite{horn1981finite,orland1990lattice,chandrasekharan1997quantum}, which allow the exact embodiment of non-Abelian gauge interactions in ultracold matter. On the other hand, we utilize fundamental symmetries of matter, such as the $SU(2I+1)$ invariance of interactions between fermionic AE isotopes such as $^{87}$Sr or $^{173}$Yb \cite{cazalilla2009ultracold,hermele2009mott,gorshkov2010two,fukuhara2007degenerate,desalvo2010degenerate,stellmer2009bose,stellmer2011detection,sugawa2011interaction,swallows2011suppression,taie20126}. While still being far from a quantum simulator for full QCD, simpler model systems share several qualitative features, including confinement, chiral symmetry breaking ($\chi$SB), and its restoration ($\chi$SR) \cite{rajagopal2000condensed}. They provide a unique environment to investigate important dynamical questions which are out of reach for classical simulation.

\begin{figure}[tbp]
\includegraphics[width=0.5\textwidth]{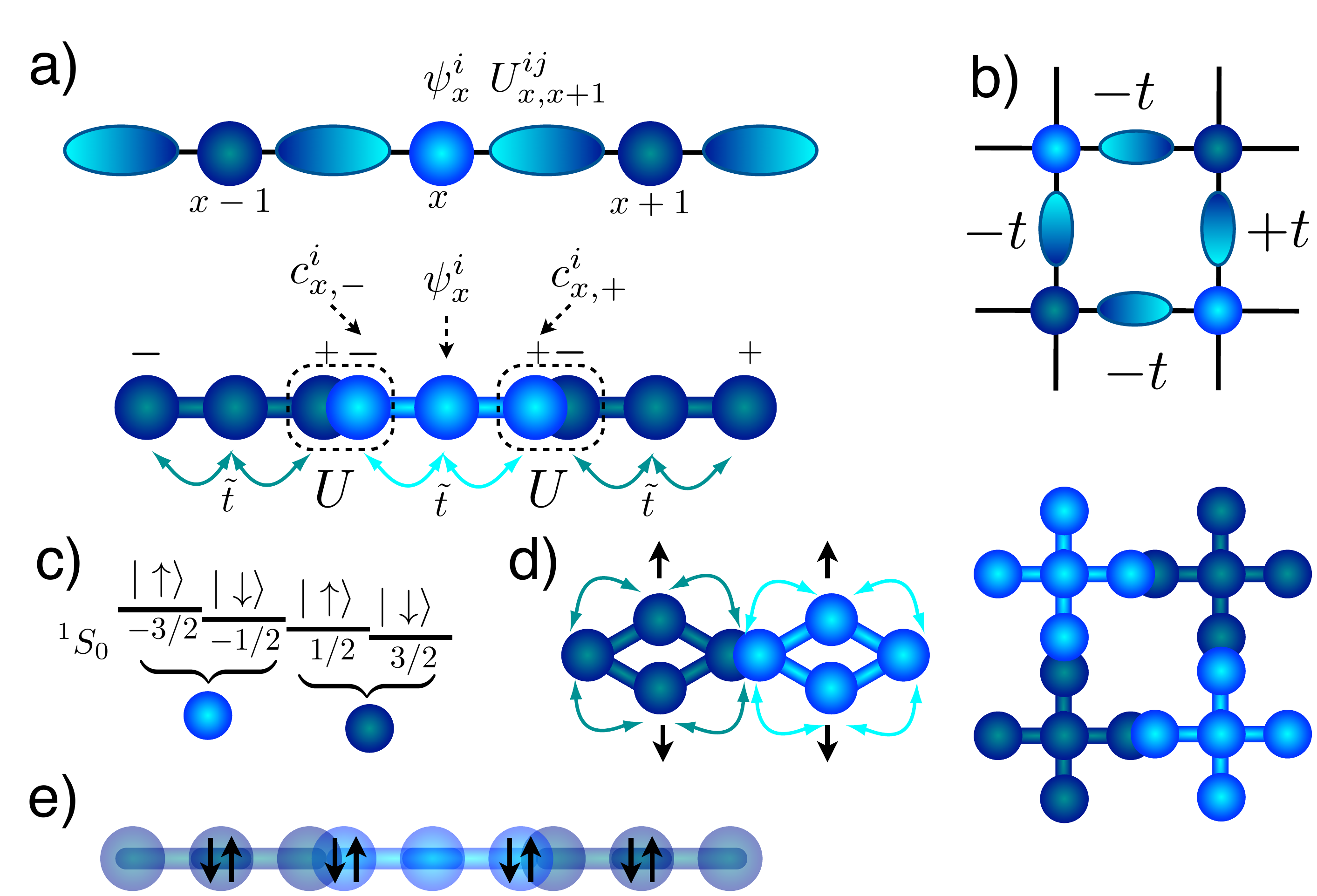}
\caption{[Color online] {\ a) (upper panel) $U(N)$ QLM in $(1+1)$D with quark fields $\protect\psi_{x}^{i}$ on lattice sites and gauge fields $U^{ij}_{x,x+1}$ on links; (lower panel) hopping of AE atoms between quark and rishon sites of the same shading. b) Implementation of the QLM in rishon representation with fermionic atoms in $(2+1)$D. c) Encoding of the color degrees of freedom for $N = 2$ ($\uparrow$, $\downarrow$) in Zeeman states of a fermionic AE atom with $I=3/2$. d) Lattice structure to avoid the interaction in fermionic matter sites using a species-dependent optical lattice (for an alternative method using site-dependent optical Feshbach resonances see the main text). e) Initial state loaded in the optical lattice with a staggered distribution of doubly occupied sites for a $U(2)$ QLM with ${\cal N} = 2$.}}
\label{QLM}
\end{figure}

The non-perturbative physics of non-Abelian gauge theories is traditionally addressed in the context of Wilson's lattice gauge theory \cite{wilson1974confinement}, in which the gluon field is represented by parallel transporter matrices residing on the links connecting neighboring lattice points of a 4D space-time lattice. Since Wilson's classical link variables take values in the continuous gauge group $SU(N)$, the corresponding Hilbert space is infinite-dimensional even for a single link. The elements of a quantum link matrix are non-commuting operators, similar to the components of a quantum spin. As a result, QLMs have a finite-dimensional Hilbert space, and therefore provide an attractive framework for the construction of quantum simulators for dynamical Abelian and non-Abelian gauge theories. In the continuum limit of QLMs, which is naturally realized via dimensional reduction, one recovers QCD with chiral quarks as domain wall fermions \cite{brower1999qcd,brower2004d}. A pedagogical introduction to QLMs, together with an extensive explanation of the corresponding terminology, is contained in the supplementary information (SI). 

\textit{Quantum Link Models.} The hopping of electrons between lattice sites $x$ and $y$ in an external magnetic background field $\vec B = \vec \nabla \times \vec A$ is described by $\psi^\dagger_x u_{xy} \psi_y$, where $u_{xy} = \exp(i \int_x^y d\vec l \cdot \vec A) \in U(1)$ is the phase picked up in this process \cite{dalibard2011colloquium}. In particle physics, gauge fields appear as dynamical quantum degrees of freedom, not just as classical background fields. Here we consider $U(N)$ and $SU(N)$ lattice gauge theories without approaching the continuum limit, using so-called staggered fermions, which are represented by creation and annihilation operators $\psi_{x}^{i \dagger}$ and $\psi_{x}^{i}$, that obey standard anti-commutation relations. Here $i \in\{1,2,\dots,N\}$ represents the non-Abelian color index of a quark. The fundamental gauge degrees of freedom representing the gluon field are $N \times N$ matrices $U_{xy}$ (with elements $U^{ij}_{xy}$) associated with the link between nearest-neighbor points $x$ and $y$ (c.f.\ Fig.\ref{QLM}a). The hopping of a quark, which exchanges color with the gluon field, is then described by $\psi^\dagger_x U_{xy} \psi_y = \psi^{i \dagger}_x U^{ij}_{xy} \psi^j_y$. This term is invariant against gauge transformations, $^\Omega \psi_x = \Omega_x \psi_x$, $^\Omega \psi_x^\dagger = \psi^{\dagger}_x \Omega^\dagger_x, ^\Omega U_{xy} = \Omega_x U_{xy} \Omega^\dagger_y$, with $\Omega_x \in U(N)$. The $SU(N)$ gauge transformations and the additional $U(1)$ gauge transformation contained in $U(N)$ are generated by 
\begin{align}
& G^{a}_{x} = \psi_{x}^{i \dagger} \lambda^{a}_{ij} \psi_{x}^{j} +
\sum_{k} \left( L^{a}_{x,x+\hat k} + R^{a}_{x-\hat k,x}\right), \nonumber \\
& G_x = \psi_{x}^{i \dagger} \psi_{x}^{j} - 
\sum_{k} \left( E_{x,x+\hat k} - E_{x-\hat k,x}\right),
\end{align} 
where $\hat k$ is a unit-vector in the $k$-direction, $\lambda^{a}$ ($a \in\{1,2,\dots,N^{2} - 1\}$) are the $SU(N)$ Gell-Mann matrices, and $f_{abc}$ are the $SU(N)$ structure constants, such that $[G^{a}_{x},G^{b}_{y}] = 2 i \delta_{xy} f_{abc} G^{c}_{x}$. The operators $L^{a}_{xy}$ and $R^{a}_{xy}$ represent $SU(N)$ electric field operators associated with the left and right end of a link $\langle xy \rangle$, while $E_{xy}$ represents the Abelian $U(1)$ electric field operator. Physical states $|\Psi\rangle$ obey the $SU(N)$ Gauss law $G_{x}^{a}|\Psi\rangle= 0$, while in a $U(N)$ gauge theory also $G_{x}|\Psi\rangle = 0$. The operators $U$, $L^{a}$, $R^{a}$, and $E$ associated with the same link obey
\begin{align}
\label{linkalgebramain} 
& [L^{a},L^{b}] = 2 i f_{abc} L^{c}, \ 
[R^{a},R^{b}] = 2 i f_{abc} R^{c}, \nonumber \\
& [L^{a},R^{b}] = [E,L^{a}] = [E,R^{a}] = 0, \nonumber \\
& [L^{a},U] = - \lambda^{a} U, \ [R^{a},U] = U \lambda^{a}, \ [E,U] = U,
\end{align}
while operators associated with different links commute.

In Wilson's lattice gauge theory, $U$ is an element of the gauge group. In a $U(N)$ gauge theory, $\mbox{det}U = \exp(i \varphi) \in U(1)$ represents a $U(1)$ link variable, canonically conjugate to the electric flux operator $E = - i \partial_{\varphi}$. In an $SU(N)$ gauge theory $U \in SU(N)$ and $L^{a}$, $R^{a}$ take appropriate derivatives with respect to the matrix elements $U^{ij}$. The resulting Hilbert space per link is then unavoidably infinite-dimensional. In order to represent the commutation relations of the gauge algebra of Eq.(\ref{linkalgebramain}) in a finite-dimensional Hilbert space, QLMs give up the commutativity of the matrix elements $U^{ij}$ without compromising gauge invariance. The real and imaginary parts of the matrix elements $U^{ij}$ of the $N \times N$ quantum link matrix are represented by $2N^{2}$ Hermitean operators. Together with the electric field operators $L^{a}$, $R^{a}$, and $E$ these are $2N^{2} + 2(N^{2}-1) + 1 = (2N)^{2} - 1$ generators which form the embedding algebra $SU(2N)$. While $U(1)$ quantum links can be represented by quantum spins embedded in an $SU(2)$ algebra, $U(N)$ or $SU(N)$ QLMs can be realized with different representations of $SU(2N)$. A useful representation is based on fermionic rishon constituents \cite{brower2004d}
\begin{align}
& L^{a} = c^{i \dagger}_{+} \lambda^{a}_{ij} c^{j}_{+}, \ R^{a} = c^{i
\dagger}_{-} \lambda^{a}_{ij} c^{j}_{-}, \ E = \frac{1}{2}(c^{i \dagger}_{-}
c^{i}_{-} - c^{i \dagger}_{+} c^{i}_{+}),\nonumber\\
& U^{ij} = c^{i}_{+} c^{j \dagger}_{-}, \ \mathcal{N} = c^{i \dagger}_{-}
c^{i}_{-} + c^{i \dagger}_{+} c^{i}_{+}.
\end{align}
The rishon creation and annihilation operators, $c^{i\dagger}_{\pm}$ and $c^{i}_{\pm}$, are associated with the left and right ends of a link (c.f.\ Fig.\ref{QLM}a) and obey standard anti-commutation relations. Our construction of a quantum simulator for $U(1)$ gauge theories used Schwinger bosons to represent quantum links \cite{banerjee2012}. Here it is natural to replace Schwinger bosons by rishon fermions. $\mathcal{N}$ counts the number of rishons on a link.

The Hamiltonian of a $(d+1)$D $U(N)$ QLM with staggered fermions takes the form
\begin{align}
\label{Hamiltonian}\!\!\!\!\!H & =- t \sum_{\langle x y \rangle} \left( s_{xy}
\psi_{x}^{i \dagger} U_{xy}^{ij} \psi_{y}^{j} + \mathrm{h.c.}\right)  + m
\sum_{x} s_{x} \psi_{x}^{i \dagger} \psi_{x}^{i} \nonumber \\
& =- t \sum_{\langle x y \rangle} \left( s_{xy} Q_{x,+k}^{\dagger}Q_{y,-k} +
\mathrm{h.c.}\right)  + m \sum_{x} s_{x} M_{x},
\end{align}
where $s_{x} = (-1)^{x_{1} + \dots+ x_{d}}$ and $s_{xy} = (-1)^{x_{1} + \dots + x_{k-1}}$, with $y = x + \hat k$. $t$ is the strength of the hopping term, and $m$ is the mass. The summation convention is implicit in the color indices. We have also introduced the $U(N)$ gauge invariant ``meson'' and ``constituent quark'' operators $M_{x} = \psi_{x}^{i\dagger} \psi_{x}^{i}$ and $Q_{x,\pm k} = c^{i \dagger}_{x,\pm k} \psi^{i}_{x}$. Together with the ``glueball'' operators $\Phi_{x,\pm k,\pm l} = c^{i\dagger}_{x,\pm k} c^{i}_{x,\pm l}$, they form a site-based $U(2d+1)$ algebra. The rishon number is conserved locally on each link. The $U(N)$ model has no baryons, since the $U(1)$ baryon number symmetry is gauged. In order to obtain charge conjugation invariance $C$ and to reduce the gauge symmetry to $SU(N)$, one must work with $\mathcal{N}_{xy} = N$ rishons per link. Adding the term $\gamma \sum_{\langle xy \rangle} (\text{det}U_{xy} + \text{h.c.})$ to the Hamiltonian, explicitly breaks the $U(N)$ gauge symmetry down to a local $SU(N)$ and a global $U(1)$ baryon number symmetry generated by $B  = \sum_x \left( \psi_x^{i\dagger} \psi_x^i - \frac{N}{2} \right)$. The symmetries of various model systems are summarized in Table 1. All models have a ${\mathbb{Z}}(2)$ chiral symmetry, which is spontaneously broken at a critical temperature $T_{c}$, and may get restored at non-zero baryon density $n_{B}$.
\begin{table}[ptb]
\begin{center}
\begin{tabular}
[c]{|c|c|c|c|c|c|c|}\hline
dimension & group & $\mathcal{N}$ & $C$ & flavor & baryon &
phenomena\\\hline\hline
$(1+1)$D & $U(2)$ & 1 & no & no & no & $\chi$SB, $T_{c} = 0$
\\ \hline
$(2+1)$D & $U(2)$ & 2 & yes & ${\mathbb{Z}}(2)$ & no & $\chi$SB,
$T_{c} > 0$
\\ \hline
$(2+1)$D & $SU(2)$ & 2 & yes & ${\mathbb{Z}}(2)$ & $
\begin{array}[c]{c} U(1) \\ \mbox{boson} \end{array} $ & $
\begin{array}[c]{c} \chi\mbox{SB}, T_{c} > 0 \\ \chi\mbox{SR}, n_{B} > 0
\end{array} $
\\ \hline
$(3+1)$D & $SU(3)$ & 3 & yes & ${\mathbb{Z}}(2)^{2}$ & $
\begin{array} [c]{c} U(1) \\ \mbox{fermion} \end{array} $ & $
\begin{array}[c]{c} \chi\mbox{SB}, T_{c} > 0 \\ \chi\mbox{SR}, n_{B} > 0
\end{array} $
\\ \hline
\end{tabular}
\end{center}
\caption{Symmetries and phenomena in some QLMs.}
\end{table}
It would be natural to add electric and magnetic field energy terms $\frac{g^{2}}{2} \sum_{\langle x y \rangle} \left( L_{xy}^{a} L_{xy}^{a} + R_{xy}^{a} R_{xy}^{a}\right) $, $\frac{{g^{\prime}}^{2}}{2} \sum_{\langle x y \rangle} E_{xy}^{2}$, and $\frac{1}{4 g^{2}} \sum_{\langle w x y z \rangle} \left( U_{wx} U_{xy} U_{yz} U_{zw} + \mathrm{h.c.}\right)$, where $\langle w x y z \rangle$ denotes an elementary plaquette with $g^2$ and $g^{\prime 2}$ as the coupling constants. At strong coupling these terms are inessential for qualitative features of the dynamics at finite temperature or baryon density, and are thus not yet included in our implementation.

\textit{Atomic quantum simulation of $\mathit{U(N)}$ QLMs.} An illustration of the QLM and its rishon representation for (1+1)D and (2+1)D is provided in Fig.\ref{QLM}. Quark fields $\psi_{x}^{i}$ reside on the lattice sites $x$, while the rishons $c_{x,\pm k}^{i}$ are on ``link-sites'' $(x,\pm k)$ at the left (right) end of the links exiting (entering) the point $x$ (c.f.\ Fig.\ref{QLM}a lower panel). The key step in our physical implementation is to interpret the \emph{lattice with quark and rishon sites} in Figs.\ref{QLM}a,b as a \textit{physical optical lattice for fermionic atoms}. Hence, an atom on site $x$ of the optical lattice represents a quark $\psi_{x}^{i}$, while hopping of this atom to a link-site $(x,\pm k)$ converts it to a rishon $c_{x,\pm k}^{i}$. The color index $i$ is encoded in internal atomic states.

\begin{figure}[ptb]
\includegraphics[width=0.42\textwidth]{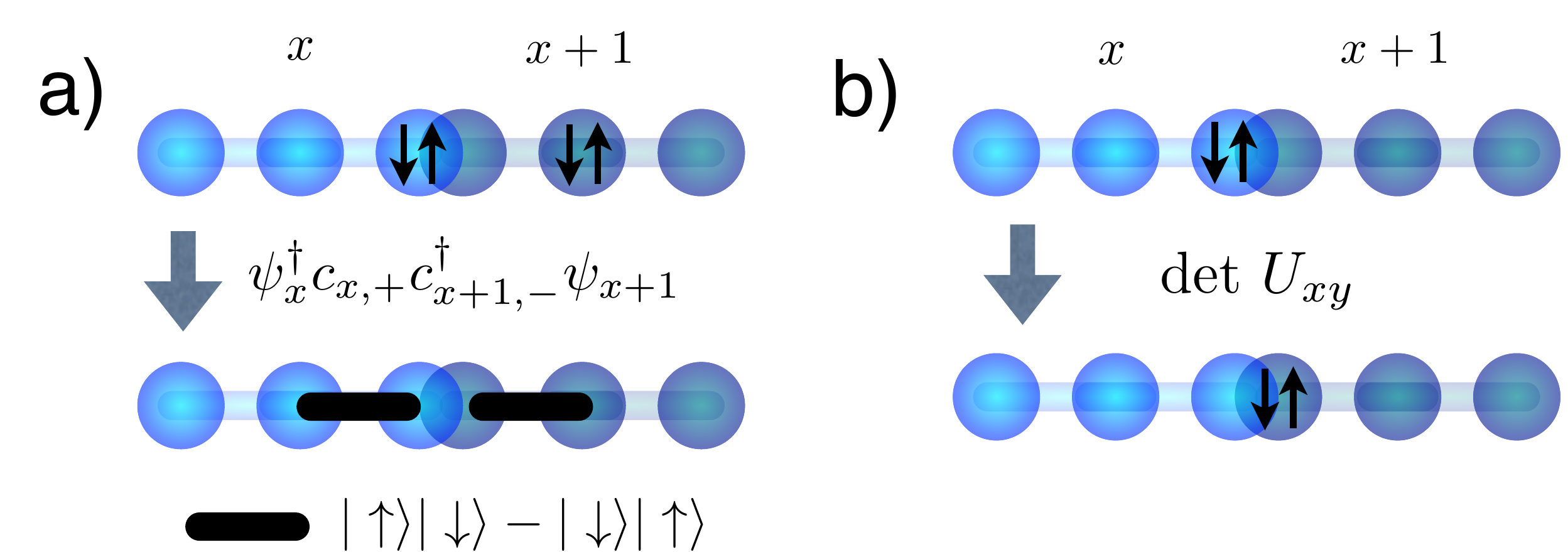}
\caption{[Color online] Dynamical processes in $U(2)$ QLMs with ${\cal N} = 2$. a) Matter-gauge interaction as correlated hopping of quarks and rishons. Starting with a configuration of site-singlets, the matter-gauge interaction converts them into nearest-neighbor singlets, keeping the rishon number per link constant. b) The determinant term corresponds to two-body hopping of both rishons on the link.}
\label{dynamics}
\end{figure}

The basic building blocks in our atomic setup are the tunnel-coupled triple-wells in (1+1)D (Fig.\ref{QLM}a) or the cross-shaped vertices in (2+1)D (Fig.\ref{QLM}b). The corresponding hopping dynamics of the atoms is described by the Hamiltonian $h_{x,k} = \tilde{t}(s_{xy} Q_{x,+k} + Q_{x,-k} + \mathrm{h.c.})$. Physically, the overlap of the Wannier wave functions can be used to implement the usual tunneling \cite{lewenstein2012ultracold}. In case different phases are needed to simulate staggered fermions in the lattice, Raman assisted tunneling \cite{dalibard2011colloquium} or shaken optical lattices \cite{lignier2007dynamical,struck2011quantum} can be applied. In order to obtain the desired quark-rishon dynamics, we introduce the microscopic atomic Hamiltonian 
\begin{equation}
\widetilde{H} = U \sum_{\langle xy \rangle}(\mathcal{N}_{xy} - n)^2
+ \sum_{x,k} h_{x,k} + m \sum_{x} s_{x} M_{x}.
\label{Htilde}
\end{equation}
The first term enforces the constraint of $\mathcal{N}_{xy} = n$ rishons per link, with $U\gg\tilde{t}$. In a physical setup, this is implemented as a strong repulsion  between atoms occupying rishon-sites, indicated in Fig.\ref{QLM}a by the overlapping link-sites, and by a potential off-sets in the rishon sites. Details on the lattice structure are discussed in the SI. The second term represents atomic hopping, while the last term realizes the staggered fermion mass with a superlattice. In second order perturbation theory in the tunnel-coupling, the above Hamiltonian induces the hopping term of Eq.(\ref{Hamiltonian}) with $t = \tilde{t}^{2}/U$. Fig.\ref{dynamics}a illustrates the matter-gauge interaction. We note that an additional term $t \sum_{x,\pm k} Q_{x,\pm k}^{\dagger} Q_{x,\pm k}$ is also generated. This is no problem, because this term is invariant under all relevant symmetries. It is also possible to add a 4-fermion term $V \sum_x M_x^2$.

With the Hamiltonian of Eq.(\ref{Htilde}) we have reduced the realization of $U(N)$ QLMs to a lattice dynamics of interacting fermions. This is enabled by the factorization of the quantum link variables into rishons. We emphasize that the building blocks in $\widetilde{H}$ are \emph{gauge invariant} ``meson'' and ``constituent quark'' operators, which allows a gauge invariant implementation of the dynamics. This is in contrast to previous work, where Gauss' law was enforced by an energy constraint in the microscopic dynamics. The essential symmetries of $\widetilde{H}$ to be respected by the implementation are: (i) the color-independent hopping of fermions and rishons, and (ii) the color-independent interaction between rishons to ensure the \emph{local} particle number conservation on each link. Indeed these symmetries are accurately respected in setups with AE atoms \cite{cazalilla2009ultracold,gorshkov2010two,taie20126}.

For a given nuclear spin $I$, the electronic ground state $^{1}S_{0}$ of fermionic AE atoms has $2I+1$ Zeeman levels $m_{I} = -I,\dots,+I$. We encode the color degrees of freedom for the even (odd) building blocks (triple-wells in (1+1)D and cross-shaped vertices in (2+1)D, represented by the light (dark) shading in Fig.\ref{QLM}) in the $N$ lowest (highest) $m_{I}$ levels (c.f.\ Fig.\ref{QLM}c). For example, to implement a $U(2)$ QLM, we choose positive nuclear spin states $m_{I} = 3/2, 1/2$ on the even and negative nuclear spin states $m_{I} = -3/2,-1/2$ on the odd building blocks. The AE atoms have the unique property that their scattering is almost exactly $SU(2I+1)$-symmetric, i.e., all pairs of states have the same scattering length \cite{cazalilla2009ultracold,gorshkov2010two,taie20126}. This guarantees the symmetry of the $U$ term in Eq.(\ref{Htilde}). The $m_I$-dependent hopping illustrated in Fig.\ref{QLM}a can be realized in optical lattices with an appropriate choice of laser frequencies and polarizations \cite{daley2008quantum,yi2008state}, or with optical potentials obtained by holographic techniques \cite{bakr2009quantum,weitenberg2011single}. Finally, the repulsion $U$, which only affects the rishon- but not the quark-sites, can be realized with optical Feshbach resonances of AE atoms allowing spatially dependent on-site interactions \cite{ciurylo2005optical,naidon2006optical,enomoto2008optical,reichenbach2009controlling,chin2010feshbach,blatt2011measurement,note1}. An alternative setup uses $m_I$-dependent optical lattices with overlapping sites for the interacting, and spatially separated sites for the non-interacting fermions (c.f.\ Fig.\ref{QLM}d).

$SU(N)$\textit{\ lattice gauge theories.} We now reduce the gauge symmetry from $U(N)$ to $SU(N)$ by activating the $\text{det}U_{xy}$ term. For definiteness, we investigate the $N = {\cal N} = 2$ case, for which $\text{det}U_{xy} = 2 c_{x,+k}^{1} c_{y,-k}^{1\dagger} c_{x,+k}^{2} c_{y,-k}^{2\dagger}$. This corresponds to two-particle tunneling between the overlapping rishon-sites. As indicated in Fig.\ref{dynamics}b we assume in our AE setup  partially overlapping rishon-sites implying a different overlap of the Wannier functions. This generates a repulsive interaction energy, which differs by $\Delta U$ between rishons on the same and on different link-sites, thus breaking the $SU(2I+1)$ symmetry. The two-particle transfer is now implemented as a Raman process with a Rabi frequency $\Omega$ and some large detuning $\delta$, so that single particle transitions are strongly suppressed, while a two-particle transfer can be an energy conserving process enabled by energy exchange between the atoms (see also Ref.~\cite{kraus2012}). The resulting coefficient of the $\text{det}U_{xy}$ term is $\gamma = 2 \Omega^{2} \Delta U/\delta^{2}$, which should be larger  than the typical temperature scale in cold atoms experiments.

\textit{Initial conditions and loading the optical lattice.} We now discuss how to load the lattice with the gauge invariant state illustrated in Fig.\ref{QLM}e. This state has local color-singlet pairs of atoms on alternating quark and rishon sites. It is an eigenstate of the Hamiltonian 
\begin{equation}
H_{\text{strong}} = m \sum_{x} s_{x} \psi^{i\dagger}_{x} \psi^{i}_{x} +
\gamma\sum_{\langle xy \rangle} \left(\text{det}U_{xy} + \text{h.c.}\right),
\end{equation}
which is induced by $\widetilde H$ in the limit $U \rightarrow \infty$. The $\text{det}U_{xy}$ term favors the state $|\uparrow\downarrow,0\rangle-|0,\uparrow\downarrow\rangle$ where $|\uparrow \downarrow,0\rangle = c^{\uparrow \dagger}_{x,+} c^{\downarrow \dagger}_{x,+} |0,0\rangle$ and $|0,\uparrow \downarrow \rangle = c^{\uparrow \dagger}_{y,-} c^{\downarrow \dagger}_{y,-} |0,0\rangle$. The preparation of the initial state requires to adiabatically ramp up the optical lattice on an ultracold cloud of atoms which are internally in a $50\%$ mixture of the states $m_I = 3/2, 1/2$. This leads to a band insulator with two atoms of positive nuclear spin on the dark-shaded sites in Fig.\ref{QLM}. Then, an on-site Raman two-body process will generate the desired state of Fig.\ref{QLM}e after a coherent transfer of the rishon population from the dark- to the light-shaded rishon-sites.

\textit{Imperfections and quality of gauge invariance.} The emergence of gauge symmetry from the microscopic model of Eq.(\ref{Htilde}) and the effect of imperfections have been addressed on small system sizes by means of exact diagonalization in the $U(2)$ case with ${\cal N}_{xy} = 2$. The $U(1)$ symmetry generated by $G_x$ has been verified by checking the condition $\langle \mathcal{N}_{xy} \rangle = 2 + \Delta_{\text{link}}(U/t,m/t)$, where $\Delta_{\text{link}}$ has to be small in order to ensure gauge invariance. As in Ref.\cite{banerjee2012}, we found that a relatively small $U$ is sufficient to accurately realize gauge invariance \cite{supmat}. The $SU(2)$ gauge symmetry, on the other hand, is potentially vulnerable to imperfections in the engineering of the tunneling rate. Fortunately, symmetric interactions emerge naturally thanks to the fundamental $SU(2I+1)$ symmetry of AE atoms. Since the number of fermions per basic building block (including sites of the same shading in Fig.\ref{QLM}) is conserved in the experiment up to exponentially small long-range tunneling processes, $G^3_x$ automatically commutes with $\widetilde H$. One still needs to check whether the expectation values of the operators $(G_x^1)^2, (G_x^2)^2$ vanish in the ground state. We have verified this numerically on small systems with up to 12 fermions (4 quark- and 8 rishon-sites) by introducing an imperfect color-dependent hopping rate $\tilde{t}_1 = (1-z) \tilde{t}_2$. We observe that gauge invariance is very well preserved even for $z \simeq 0.1$ \cite{supmat}.
Moreover, we emphasize that the low-energy properties of the system are expected to be robust even in the presence of small gauge variant terms (see for instance~\cite{foerster1980dynamical}).

\begin{figure}[tbp]
\includegraphics[width=0.43\textwidth]{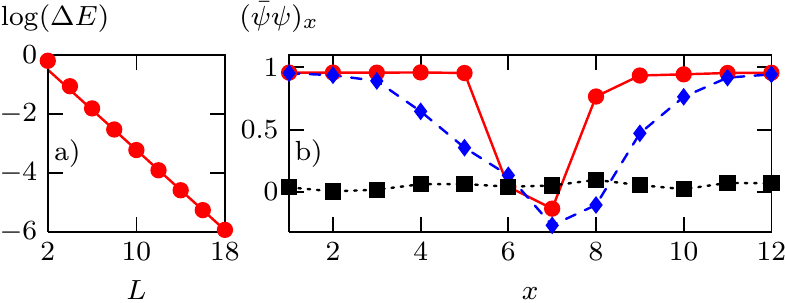}
\caption{[Color online] a) Size $L$ dependence of the energy splitting between the lowest energy eigenstates of a $U(2)$ QLM with $m=0$ and $V = - 6t$. b) Real-time evolution of the order parameter profile $(\overline\psi \psi)_x(\tau)$ for $L = 12 $, mimicking the expansion of a hot quark-gluon plasma. Here, circles (thin line), diamonds (dashed line) and squares (dotted line) correspond to $\tau/t=0, 1, 10$ respectively.}
\label{EDresults}
\end{figure}

\textit{Exact diagonalization results.} We have performed exact diagonalization studies of the $(1+1)$D $U(2)$ model with ${\cal N} = 1$ rishon per link. Figure \ref{EDresults}a shows the splitting between two almost degenerate vacuum states, which decreases exponentially with the system size $L$, thus indicating $\Z(2)$ $\chi$SB. Figure \ref{EDresults}b shows the real-time evolution of the chiral order parameter profile $(\overline\psi \psi)_x = s_x \langle \psi_x^{i \dagger} \psi_x^i - \frac{N}{2}\rangle$, starting from an initial chirally restored ``fireball'' embedded in the chirally broken vacuum. This dynamics can be observed by initializing the system in a product state of Mott double wells, and subsequently lower the lattice potential. This mimics the expanding quark-gluon plasma generated in a heavy-ion collision, and can be probed in an experimental setup by just monitoring the time-dependence of the particle density, similarly to Ref.~\cite{Trotzky2012a}.

\textit{Conclusions.} We have proposed an implementation of a quantum simulator for non-Abelian $U(N)$ and $SU(N)$ gauge theories for staggered fermions with ultracold atoms. The proposal builds on the unique properties of quantum link models with rishons representing the gauge fields: this allows a formulation in terms of a Fermi-Hubbard model, which can be realized with multi-component alkaline-earth atoms in optical lattices, and where atomic physics provides both the control fields and measurement tools for studying the equilibrium and non-equilibrium dynamics and spectroscopy. Extending such investigations towards QCD requires the incorporation of multi-component Dirac fermions with the appropriate chiral symmetries, and of additional link and plaquette terms for electric and magnetic field energies \cite{buchler2005atomic}.

\textit{Acknowledgments.} We thank P. S. Julienne, B.\ Pasquiou, and F.\ Schreck for discussions. Work at Bern is supported by the Schweizerischer Nationalfonds. Work at Innsbruck is supported by the integrated project AQUTE, the Austrian Science Fund through SFB F40 FOQUS, and by the DARPA OLE program. Authors are listed in alphabetical order.

{\it Note added}: Recently, two related works on SU(2) gauge theories in cold atomic systems have appeared~\cite{tagliacozzo2012,zohar2012}.

\bibliography{20130324citations}

\begin{thebibliography}{44}
\bibitem [{not()}]{note1}%
  \BibitemOpen
  \href@noop {} {\bibinfo  {journal} {Notice that a finite interaction of order
  $t$ in the Fermionic site will not affect gauge invariance and the form of
  the effective Hamiltonian.}\ }\BibitemShut {NoStop}%
\end{thebibliography}

\newpage

\section{Supplementary material I: $U(N)$ and $SU(N)$ Quantum Link Models with Staggered Fermions}

\subsection{Introduction}

Non-Abelian gauge fields are the key to understanding the weak and strong 
interactions between leptons and quarks, mediated by W- or Z-bosons and gluons,
respectively. The strong interactions are described by Quantum Chromodynamics 
(QCD), the relativistic quantum field theory of quarks and gluons with an 
$SU(3)$ non-Abelian gauge symmetry. Due to asymptotic freedom, at very high
energies quarks and gluons interact only weakly. At low energies, on the other
hand, they interact so strongly that they do not appear as physical states in 
the spectrum. Instead quarks and gluons are permanently bound together and thus 
confined inside hadrons. Protons and neutrons mostly contain 
up and down quarks as well as gluons, while the heavier strange quarks are more 
suppressed. Since the mass difference between up and down quarks is small 
compared to the intrinsic energy scale of QCD, there is an approximate isospin 
flavor symmetry $SU(2)_F$. Protons and neutrons form an isospin $SU(2)_F$ 
doublet. In addition, the number of quarks is conserved which gives rise to a 
$U(1)_B$ baryon number symmetry. There are different categories of hadrons. 
States with a non-vanishing baryon number, i.e.\ with a different number of 
quarks and anti-quarks, are known as baryons. In a simple-minded quark model, 
protons and neutrons are baryons consisting of three quarks of different 
$SU(3)$ colors, while mesons are quark-anti-quark pairs. The lightest mesons of
QCD are the three pions $\pi^+$, $\pi^0$, and $\pi^-$, which form an isospin 
$SU(2)_F$ triplet. The pions are the pseudo-Nambu-Goldstone bosons of the 
approximate $SU(2)_L \times SU(2)_R$ chiral symmetry of two flavor QCD, which 
gets spontaneously broken down to the isospin symmetry $SU(2)_F$ by the strong 
gauge interactions. Flavor singlet states with baryon number zero may be 
dominated by gluons and are then denoted as glueballs. 

Since quarks and gluons are confined inside hadrons (i.e.\ baryons, mesons, or 
glueballs) their low-energy dynamics cannot be studied in perturbation theory.
In order to define QCD beyond perturbation theory, Wilson has regularized the
theory on a 4-dimensional space-time lattice \cite{wilson1974confinement}. The quark fields 
then reside on
the lattice points, while the gluon fields are naturally associated with the 
links connecting neighboring points. In lattice QCD the gluon field is 
represented by $SU(3)$ matrices, which act as parallel transporters. Due to its 
non-perturbative nature and great complexity, lattice QCD requires large-scale 
Monte Carlo simulations. In this way, during the past decade very impressive 
progress has been made in determining the hadron spectrum as well as numerous
static properties of hadrons. Also the nature of the transition that separates
the low-temperature chirally broken phase from the quark-gluon plasma at high
temperatures has been understood in some detail. However, despite of these 
successes, there are also important problems in QCD that cannot be addressed 
with Monte Carlo methods. Such problems include QCD at large baryon density, 
which is relevant in nuclear and astrophysics, in particular, for understanding 
the state of matter inside neutron stars, as well as all problems concerning the
real-time dynamics of strongly interacting matter, in particular, heavy-ion
collisions. In these cases, severe sign problems prevent importance sampling
which renders the Monte Carlo method inapplicable.

Similar problems arise in strongly correlated systems in condensed matter 
physics. In recent years, several such systems have been emulated with 
ultra-cold matter in optical lattices, thus realizing an analog quantum 
simulator. Since these systems realize quantum dynamics in their hardware, they
do not suffer from the sign problem, and are thus ideally suited for 
investigating real-time dynamics. Recently, quantum simulator constructions 
have been proposed for Abelian $U(1)$ gauge theories with and without fermionic
matter. Some of these constructions take advantage of the quantum link 
formulation of lattice gauge theory, which has a finite-dimensional Hilbert
space per link. In Wilson's formulation, on the other hand, the Hilbert space is
infinite-dimensional, and thus much harder to realize in ultra-cold matter. A
non-Abelian quantum link model with gauge group $SU(2)$ has been formulated
in \cite{horn1981finite,orland1990lattice,chandrasekharan1997quantum}. Subsequently, in \cite{brower1999qcd,brower2004d} 
quantum link models have been constructed for the non-Abelian gauge groups 
$SU(N)$, $U(N)$, $SO(N)$, and $Sp(N)$, as well as for the exceptional group
$G(2)$. In particular, lattice QCD has been formulated as an $SU(3)$ quantum
link model \cite{brower1999qcd}. 
In order to reach the continuum limit, the model is formulated with
an additional dimension of finite extent, which ultimately disappears via
dimensional reduction. In this framework, quarks with a chiral symmetry emerge 
naturally as domain wall fermions. While it may take a long time until a
quantum simulator will be realized for quantum link QCD in the continuum limit,
it is timely to investigate simpler non-Abelian gauge theories in order to
address non-perturbative questions that are inaccessible to classical 
simulation.

In this paper, we construct quantum simulators for $U(N)$ and $SU(N)$ lattice 
gauge theories in $(1+1)$ and $(2+1)$ space-time dimensions coupled to
staggered fermions with a $\Z(2)$ chiral symmetry. Being discrete, this chiral
symmetry can break spontaneously in just one spatial dimension, at least at
zero temperature. In two spatial dimensions it can break spontaneously even at 
finite temperature. In $(2+1)$ dimensions, staggered fermions have an additional
discrete $\Z(2)_F$ flavor symmetry, which allows us to distinguish ``mesons''
from ``glueballs''. In a $U(N)$ gauge theory, baryon number is part of the local
symmetry. Hence, baryons violate Gauss' law and thus do not belong to the
physical spectrum. In an $SU(N)$ gauge theory, on the other hand, there is a
global $U(1)_B$ symmetry that distinguishes physical states with different
baryon numbers. In an $SU(2)$ gauge theory two quarks are confined inside a 
baryon. Then, in contrast to the real world, baryons are bosons. Despite 
numerous differences with QCD, concerning the space-time dimension, the gauge
group, and the nature of the chiral symmetry, $U(2)$ and $SU(2)$ lattice 
gauge theories in $(1+1)$ and $(2+1)$ space-time dimensions are still useful
toy models for investigating interesting non-perturbative dynamics. For
example, one can study the real-time dynamics of chiral symmetry breaking across
a phase transition, the superfluidity of baryons resulting from the spontaneous
breakdown of $U(1)_B$, and one can even mimic heavy ion collisions.

A distinguishing feature of quantum link models is that the gauge fields can be
realized as fermion bilinears. The fermionic constituents of gauge fields are
known as rishons. As we will see, the rishon formulation is crucial for our
construction of a quantum simulator for non-Abelian gauge theories. In fact,
both ``quarks'' and rishons are represented by the same fermionic atoms that
hop between the sites of an optical lattice. From a theoretical perspective,
the color indices of ``quarks'' and rishons can be contracted to form 
color-neutral ``meson'', ``glueball'', and ``constituent quark'' operators. It 
is remarkable that the theory can be formulated entirely in terms of these 
$U(N)$ or $SU(N)$ gauge invariant objects. Thanks to the explicit reduction to
the gauge invariant sector, exact diagonalization studies are possible even on
moderate size systems, which can be used to validate an experimentally realized
quantum simulator. Since the number of rishons per link is conserved locally, 
in the $U(N)$ or $SU(N)$ gauge invariant formulation of the theory, an 
additional $U(1)$ link gauge invariance emerges. This ensures that the confining
dynamics of ``quarks'' and ``gluons'' is still correctly represented by
``mesons'', ``glueballs'', and ``constituent quarks''.

\begin{figure}[htb]
\epsfxsize=70mm
\epsffile{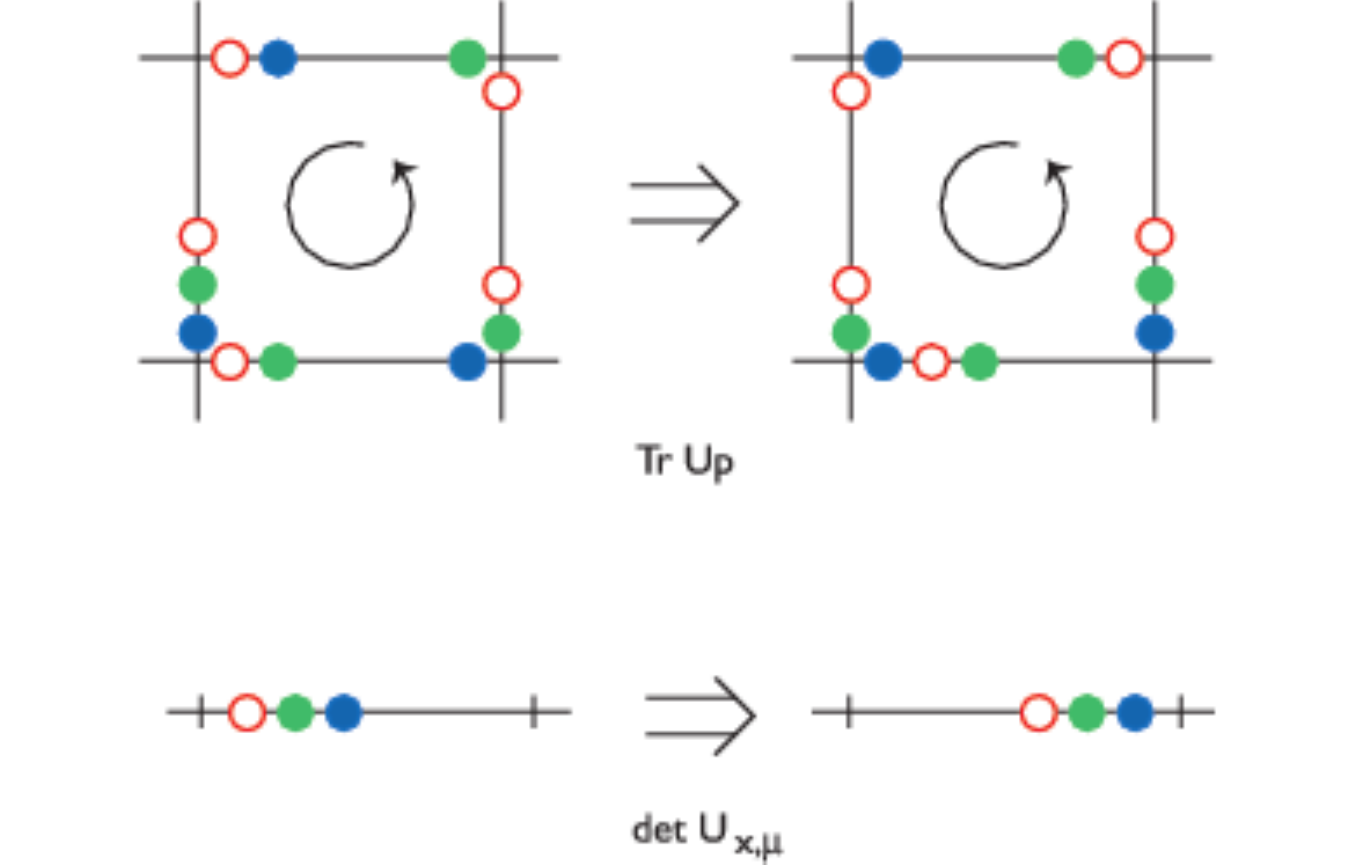}
\caption{Rishon dynamics: The trace part of the Hamiltonian induces a hopping of
rishons of various colors around a plaquette. The determinant part shifts a
color-neutral rishon baryon from one end of a link to the other.}
\end{figure}

\subsection{$U(N)$ and $SU(N)$ Quantum Link Models}

Let us consider a system of staggerd fermions on a $d$-dimensional spatial
lattice coupled to a $U(N)$ gauge field. The corresponding Hamiltonian then
takes the form
\begin{eqnarray}
\label{Hamiltonian}
H&=&- t \sum_{\langle x y \rangle} 
\left(s_{xy} \psi_x U_{xy} \psi_y + \mathrm{h.c.}\right) + 
m \sum_x s_x \psi_x^{i \dagger} \psi_x^i \nonumber \\
&+&\frac{g^2}{2} \sum_{\langle x y \rangle} 
\left(L_{xy}^a L_{xy}^a + R_{xy}^a R_{xy}^a\right) +
\frac{{g'}^2}{2} \sum_{\langle x y \rangle} E_{xy}^2 \nonumber \\
&-&\frac{1}{4 g^2}
\sum_{\langle w x y z \rangle} \left(\mathrm{Tr} U_{wx} U_{xy} U_{yz} U_{zw} + 
\mathrm{h.c.}\right) \nonumber \\
&-& \gamma \sum_{\langle x y \rangle} 
\left(\mathrm{det} U_{xy} + \mathrm{h.c.}\right).
\end{eqnarray}
Here $\psi^{i \dagger}_x$ and $\psi^i_x$ ($i \in \{1,2,\dots,N\}$ are fermion 
creation and annihilation operators that obey standard anti-commutation 
relations, $t$ is a hopping parameter, and $s_x$, $s_{xy}$ are sign-factors 
associated with the points $x$ and with the links connecting neighboring 
lattice points $x$ and $y$, respectively. The site factor is given by 
$s_x = (-1)^{x_1 + \dots + x_d}$. For the links in the 1-direction $s_{xy} = 1$, for 
those in the 2-direction $s_{xy} = (-1)^{x_1}$, and for the links in the 
$k$-direction $s_{xy} = (-1)^{x_1 + \dots + x_{k-1}}$. $U_{xy}$ is an $N \times N$ 
matrix (with matrix elements $U_{xy}^{ij}$) associated with the oriented link 
connecting the neighboring points $x$ and $y$, $E_{xy}$, $L_{xy}^a$, and 
$R_{xy}^a$ are Abelian and non-Abelian electric field operators associated with 
the same link, and $g'$ and $g$ are the corresponding Abelian and non-Abelian 
gauge couplings. The magnetic field energy is represented by the term 
associated with the elementary plaquettes $\langle w x y z \rangle$. Finally, 
the term proportional to $\gamma$ explicitly breaks a $U(N)$ gauge symmetry 
down to $SU(N)$.

The various operators obey the commutation relations
\begin{eqnarray}
\label{linkalgebra}
&&[L^a,L^b] = 2 i f_{abc} L^c, \ [R^a,R^b] = 2 i f_{abc} R^c, \nonumber \\
&&[L^a,R^b] = [E,L^a] = [E,R^a] = 0, \nonumber \\
&&[L^a,U] = - \lambda^a U, \ [R^a,U] = U \lambda^a, \ [E,U] = U.
\end{eqnarray}
Here we have suppressed the link index $xy$. Operators associated with different
links commute with each other. The Hermitean generators of $SU(N)$ obey the
commutation relations
\begin{equation}
[\lambda^a,\lambda^b] = 2 i f_{abc} \lambda^c, \ 
\mbox{Tr} \lambda^a \lambda^b = 2 \delta^{ab},
\end{equation}
where $f_{abc}$ are the structure constants of the $SU(N)$ algebra. By 
construction, the Hamiltonian of Eq.(\ref{Hamiltonian}) is gauge invariant, 
i.e.\ it commutes with the infinitesimal generators of $SU(N)$ gauge 
transformations
\begin{eqnarray}
&&G_x^a = \psi_x^\dagger \lambda^a_{ij} \psi_x^j + 
\sum_k \left(L_{x,x+\hat k} + R_{x-\hat k,x}\right), \nonumber \\
&&[G^a_x,G^b_y] = 2 i \delta_{xy} f_{abc} G^c_x,
\end{eqnarray}
where $\hat k$ is a unit-vector in the $k$-direction. A general $SU(N)$ gauge 
transformation is represented by the unitary transformation 
$V = \prod_x \exp(i \alpha_x^a G_x^a)$, which acts as
\begin{equation}
U'_{xy} = V^\dagger U_{xy} V = 
\exp(i \alpha_x^a \lambda^a) U_{xy} \exp(- i \alpha_y^a \lambda^a).
\end{equation}
For $\gamma = 0$ the Hamiltonian has an additional $U(1)$ gauge symmetry which 
is generated by
\begin{eqnarray}
&& G_x = \psi_x^{i \dagger} \psi_x^i - 
\sum_k \left(E_{x,x+\hat k} - E_{x-\hat k,x}\right), \nonumber \\
&& [G_x,G_x^a] = 0.
\end{eqnarray}
A general $U(1)$ gauge transformation, represented by the unitary 
transformation $W = \prod_x \exp(i \alpha_x G_x)$, acts as
\begin{equation}
U'_{xy} = W^\dagger U_{xy} W = \exp(i \alpha_x) U_{xy} \exp(- i \alpha_y).
\end{equation}

In Wilson's formulation of lattice gauge theory, the link matrices $U_{xy}$ are 
parallel transporters that take values in the gauge group $U(N)$ or $SU(N)$, and
$E_{xy}$, $L_{xy}^a$, and $R_{xy}^a$ are the corresponding canonically conjugate 
momentum operators, which take derivatives with respect to the matrix elements 
of $U_{xy}$. In that case, the commutation relations of Eq.(\ref{linkalgebra})
are realized in an infinite-dimensional Hilbert space per link, which poses
great challenges for the implementation in a quantum simulator. Quantum link
models extend the theoretical framework of lattice gauge theories by realizing
the algebra of Eq.(\ref{linkalgebra}) in a finite-dimensional Hilbert space per
link. This is achieved by giving up the classical commutative nature of the 
matrix elements of $U_{xy}$. In contrast to Wilson's lattice gauge theory, in
quantum link models the real and imaginary parts of $U_{xy}^{ij}$ are no longer
represented by real numbers, but by non-commuting Hermitean operators. In this
sense, quantum links are gauge covariant generalizations of quantum spins. 

On each link the commutation relations of Eq.(\ref{linkalgebra}) can be 
realized by using the generators of an $SU(2N)$ algebra. In particular, the 
real and imaginary parts of the $N^2$ matrix elements $U_{xy}^{ij}$ are 
represented by $2N^2$ Hermitean generators of the embedding algebra $SU(2N)$.
The $2(N^2 - 1)$ generators $L^a$ and $R^a$ of $SU(N)_L \times SU(N)_R$ gauge
transformations on the left and on the right end of the link also belong to the
$SU(2N)$ algebra. The Abelian $U(1)$ gauge transformations do not distinguish 
between left and right and are represented by the additional generator $E$.
Altogether there are thus $2 N^2 + 2 (N^2 - 1) + 1 = 4 N^2 - 1$ generators,
which form the algebra $SU(2N)$. This requires that, unlike in Wilson's lattice
gauge theory, the elements of a link matrix, $U^{ij}$, are non-commuting 
operators that obey the commutation relations
\begin{eqnarray}
&& [U^{ij},U^{kl\dagger}] = 
2 (\delta_{ik} \sigma^{a *}_{jl} R^a - \delta_{jl} \sigma^a_{ik} L^a +
2 \delta_{ik} \delta_{jl} E), \nonumber \\
&& [U_{ij},U_{kl}] = [(U^\dagger)_{ij},(U^\dagger)_{kl}] = 0.
\end{eqnarray}
Again, we should emphasize that the commutation relations are local, namely all 
commutators between operators assigned to different links are zero. 

Since all elements of $U_{xy}$ commute with each other (although $U_{xy}$ and
$U^\dagger_{xy}$ do not commute) the definition of $\mbox{det} U_{xy}$ does not 
suffer from operator ordering ambiguities. By construction, for $\gamma \neq 0$,
the Hamiltonian of Eq.(\ref{Hamiltonian}) is invariant only under $SU(N)$ but 
not under additional $U(1)$ gauge transformations. However, there is a subtlety 
that needs to be taken into account. In the fundamental representation of 
$SU(2N)$, the operator that represents $\mbox{det} U_{xy}$ turns out to vanish. 
Hence, in that case even the Hamiltonian of Eq.(\ref{Hamiltonian}) has a $U(N)$ 
gauge invariance. On the other hand, since the above commutation relations can 
be realized with any representation of $SU(2N)$, we can use higher 
representations in which the determinant in general does not vanish. The 
$(2N)!/(N!)^2$-dimensional representation of $SU(2N)$ is the smallest with a 
non-vanishing determinant. For an $SU(2)$ gauge theory the Hilbert space of the
corresponding quantum link model then has 6 states per link, while for $SU(3)$
it has 20 states per link. In the standard Wilson formulation of lattice gauge 
theory, on the other hand, the single link Hilbert space is already
infinite-dimensional.

\subsection{Global Symmetries of the $U(N)$ and $SU(N)$ Quantum Link Model}

Let us discuss the global symmetries of the Hamiltonian of 
Eq.(\ref{Hamiltonian}). First, we consider the shift symmetry $S_k$ by one 
lattice spacing in the $k$-direction
\begin{eqnarray}
\label{S}
&&^{S_k} \psi^i_x = (-1)^{x_{k+1}+ \dots + x_d} \psi^i_{x+\hat k}, \ 
^{S_k} U^{ij}_{xy} = U^{ij}_{x+\hat k,y+\hat k}, \nonumber \\
&&^{S_k} L^a_{xy} = L^a_{x+\hat k,y+\hat k}, \ 
^{S_k} R^a_{xy} = R^a_{x+\hat k,y+\hat k}, \nonumber \\
&&^{S_k} E^a_{xy} = E^a_{x+\hat k,y+\hat k}.
\end{eqnarray}
Only the mass term breaks this symmetry explicitly. Shifts by an even number 
of lattice spacings in the $k$-direction play the role of ordinary translations.
The associated conserved quantum number is the momentum 
$p_k \in ]-\frac{\pi}{2a},\frac{\pi}{2a}]$. In the zero-momentum sector the
different shifts $S_k$ form the group $\Z(2)^d$. Simultaneous shifts by one 
lattice spacing in an even number of different directions play the role of 
flavor transformations, which form the subgroup $\Z(2)^{d-1}$. Simultaneous 
shifts by one lattice spacing in an odd number of different directions can be
decomposed into a flavor transformation and a $\Z(2)$ chiral transformation,
which is explicitly broken by the staggered fermion mass term. In one spatial
dimension, the chiral transformation is simply the shift by one lattice spacing.
In two dimensions, simultaneous shifts by one lattice spacing in both directions
represent a $\Z(2)$ flavor transformation, while a shift in either the 1- or the
2-direction correspond to the $\Z(2)$ chiral symmetry.

In a $U(N)$ gauge theory, there are no baryons because the $U(1)$ symmetry 
contained in $U(N)$ is gauged. Hence, baryons (which are ``charged'' under this
$U(1)$ symmetry) violate the Gauss law and are thus excluded from the physical
Hilbert space. In an $SU(N)$ gauge theory, on the other hand, there is a global
$U(1)_B$ baryon number symmetry, i.e.\ $[H,B] = 0$, which is generated by
\begin{equation}
B = \sum_x \left(\psi^{i \dagger}_x \psi^i_x - \frac{N}{2}\right).
\end{equation}
Note that baryon number is counted relative to a filled Dirac sea.

Another important symmetry is charge conjugation, which for staggered fermions
also involves a shift by one lattice spacing. As for the chiral and flavor
symmetries, there are separate symmetries $C_k$ for the different directions.
\begin{eqnarray}
\label{C}
&&^{C_k} \psi^i_x = (-1)^{x_1 + \dots + x_k} \psi^{i \dagger}_{x+\hat k}, \
^{C_k} U^{ij}_{xy} = U^{ij \dagger}_{x+\hat k,y+\hat k}, \nonumber \\
&&^{C_k} L^a_{xy} = - L^{a *}_{x+\hat k,y+\hat k}, \ 
^{C_k} R^a_{xy} = - R^{a *}_{x+\hat k,y+\hat k}, \nonumber \\
&&^{C_k} E^a_{xy} = - E^a_{x+\hat k,y+\hat k}.
\end{eqnarray}
A combination of the two symmetries $C_k$ and $C_l$ acts as
\begin{equation}
^{C_l C_k} \psi^i_x = - ^{S_l S_k} \psi^i_x.
\end{equation}
This implies that, up to an overall minus-sign, the combination of $C_k$ and
$C_l$ corresponds to a flavor transformation. A combination of an odd number of
$C_k$ operations, on the other hand, plays the role of a genuine charge 
conjugation. It should be noted that, strictly speaking, charge conjugation is
not a $\Z(2)$ symmetry, because applying it twice is equivalent to a 
translation, up to a minus-sign. Charge conjugation is a symmetry only if one
chooses a non-chiral (i.e.\ real or pseudo-real) representation of the
embedding algebra $SU(2N)$. Finally, let us also list the parity symmetry $P$, 
which acts as
\begin{eqnarray}
\label{P}
&&^P \psi^i_x = \psi^i_{-x}, \ ^P U^{ij}_{xy} = U^{ji \dagger}_{-y,-x}, \nonumber \\
&&^P L^a_{xy} = R^a_{-y,-x}, \ ^P R^a_{xy} = L^a_{-y,-x},   \nonumber \\
&& ^PE_{xy} = - E_{-y,-x}.
\end{eqnarray}

\subsection{The Rishon Representation of Quantum Links}

In this section we formulate quantum link models using anti-commuting 
operators describing fermionic constituents of the gluons. The algebraic 
structure of a quantum link model is determined by the commutation relations 
derived above. The Hilbert space is a direct product of representations of 
$SU(2N)$ on each link, with the generators on different links commuting with 
each other. We can thus limit ourselves to a single link. The commutation 
relations can be realized in a representation using anti-commuting rishon 
operators $c^i_{x,\pm k}$, $c^{i \dagger}_{x,\pm k}$ with color index 
$i \in \{1,2,...,N\}$. The rishon operators are associated with the left and 
right ends of a link and are characterized by a lattice point $x$ and a link 
direction $\pm k$. They obey canonical anti-commutation relations
\begin{eqnarray}
&& \{c^i_{x,\pm k},c^{j \dagger}_{y,\pm l}\} = \delta_{xy} \delta_{\pm k,\pm l} \delta_{ij},  \nonumber \\
&& \{c^i_{x,\pm k},c^j_{y,\pm l}\} = \{c^{i \dagger}_{x,\pm k},c^{j \dagger}_{y,\pm l}\} = 0.
\end{eqnarray}
It should be noted that quark and rishon operators also anti-commute with each
other. Under $SU(N)$ gauge transformations the operators $c$ and $c^\dagger$ 
transform in the fundamental and anti-fundamental representation, respectively. 
It is straightforward to show that the commutation relations of 
Eq.(\ref{linkalgebra}) are satisfied when we write 
\begin{eqnarray}
&& L^a_{xy} = c^{i \dagger}_{x,+} \lambda^a_{ij} c^j_{x,+}, \
R^a_{xy} = c^{i \dagger}_{y,-} \lambda^a_{ij} c^j_{y,-}, \nonumber \\
&& E_{xy} = \frac{1}{2}(c^{i \dagger}_{y,-} c^i_{y,-} - c^{i \dagger}_{x,+} c^i_{x,+}), \
U^{ij}_{x,y} = c^i_{x,+} c^{j \dagger}_{y,-}.
\end{eqnarray}
All operators introduced so far (including the Hamiltonian) commute with the 
rishon number operator
\begin{equation}
{\cal N}_{xy} = c^{i \dagger}_{y,-} c^i_{y,-} + c^{i \dagger}_{x,+} c^i_{x,+}
\end{equation}
on each individual link. Hence, we can limit ourselves to superselection 
sectors of fixed rishon number for each link. This is equivalent to working in 
a given irreducible representation of $SU(2N)$. Let us use the rishon 
representation to take a closer look at the determinant operator that we used 
to break the $U(N)$ gauge symmetry down to $SU(N)$. We have
\begin{eqnarray}
\mbox{det} U_{xy}&=&\frac{1}{N!} \epsilon_{i_1 i_2 ... i_N} 
U_{xy}^{i_1 i'_1} U_{xy}^{i_2 i'_2} \dots U_{xy}^{i_N i'_N} \epsilon_{i'_1 i'_2 \dots i'_N}
\nonumber \\ 
&=&\frac{1}{N!} \epsilon_{i_1 i_2 ... i_N} c^{i_1}_{x,+} c^{i'_1 \dagger}_{y,-} 
c^{i_2}_{x,+} c^{i'_2 \dagger}_{y,-} \dots c^{i_N}_{x,+} c^{i'_N \dagger}_{y,-} 
\epsilon_{i'_1 i'_2 \dots i'_N} \nonumber \\
&=&N! \ c^1_{x,+} c^{1 \dagger}_{y,-} c^2_{x,+} c^{2 \dagger}_{y,-} \dots 
c^N_{x,+} c^{N \dagger}_{y,-}.
\end{eqnarray}
Only when this operator acts on a state with exactly ${\cal N} = N$ rishons
(all of a different color), it can give a non-zero contribution. In all other 
cases the determinant vanishes. This means that we can reduce the symmetry from
$U(N)$ to $SU(N)$ via the determinant only when we work with exactly 
${\cal N} = N$ fermionic rishons on each link. The number of fermion states per
link then is
\begin{equation}
\left( \begin{array}{c} 2N \\ N \end{array} \right) = \frac{(2N)!}{(N!)^2}.
\end{equation}
This is the dimension of the $SU(2N)$ representation with a totally 
antisymmetric Young tableau with $N$ boxes (arranged in a single column).

The various symmetries act on the rishons as
\begin{eqnarray}
\label{rishonsym}
^{S_k} c^i_{x,\pm l} = c^i_{x+\hat k,\pm l}, \
^{C_k} c^i_{x,+l} = c^{i \dagger}_{x+\hat k,+l}, \nonumber \\
^{C_k} c^i_{x,-l} = - c^{i \dagger}_{x+\hat k,-l}, \
^P c^i_{x,\pm l} = c^i_{-x,\mp l}. 
\end{eqnarray}
It is straightforward to show that these transformations induce the 
corresponding transformations of Eqs.(\ref{S}), (\ref{C}), and (\ref{P}).

\subsection{Glueballs, Mesons, and Constituent Quarks}

In this section we express the Hamiltonian of the $U(N)$ quantum link model 
with quarks in terms of color neutral operators. Gauge invariance requires that
these operators are local bilinear combinations of rishons and quarks, which we
refer to as glueballs, mesons, and constituent quarks. All these objects ---
including the constituent quarks --- are bosons. The determinant term in the 
$SU(N)$ quantum link Hamiltonian would give rise to an additional color-singlet 
rishon-baryon consisting of $N$ rishons. For odd $N$ this object would hence be 
a fermion. To avoid complications related to these objects we limit ourselves 
to $U(N)$ in this section.
 
By contracting the color indices of two rishons, we construct ``glueball'' 
operators
\begin{equation}
\Phi_{x,\pm k,\pm l} = c^{i \dagger}_{x,\pm k} c^i_{x,\pm l},
\end{equation}
which satisfy the local commutation relations of $U(2d)$
\begin{eqnarray}
&&[\Phi_{x,\pm k,\pm l},\Phi_{y,\pm m,\pm n}] = \nonumber \\
&&=\delta_{xy} (\delta_{\pm l,\pm m} \Phi_{x,\pm k,\pm n} 
- \delta_{\pm k,\pm n} \Phi_{x,\pm m,\pm l}).
\end{eqnarray}
By contracting the color index of two quark fields, we construct a ``meson''
operator
\begin{equation}
M_x = \psi^{i \dagger}_x \psi^i_x,
\end{equation}
which generates a $U(1)$ algebra. While glueballs and mesons are not related 
via their commutation relations, i.e.\ $[\Phi_{x,\pm k,\pm l},M_y] = 0$, they are 
both related with the constituent quark operators that one obtains by 
contracting the color indices of a quark and a rishon
\begin{equation}
Q_{x,\pm k} = c^{i \dagger}_{x,\pm k} \psi^i_x,
\end{equation}
via the commutation relations
\begin{eqnarray}
&& [\Phi_{x,\pm k,\pm l},Q_{y,\pm m}] = 
\delta_{xy} \delta_{\pm l, \pm m} Q_{x,\pm k}, \nonumber \\
&& [M_x,Q_{y,\pm k}] = - \delta_{xy} Q_{x,\pm k}.
\end{eqnarray}
Finally, the commutation relations of the constituent quark operators take the 
form
\begin{eqnarray}
&&[Q_{x,\pm k},Q^\dagger_{y,\pm l}] = 
\delta_{xy}(\Phi_{x,\pm k,\pm l} - \delta_{\pm k,\pm l} M_x), \nonumber \\
&&[Q_{x,\pm k},Q_{y,\pm l}] = [Q^\dagger_{x,\pm k},Q^\dagger_{y,\pm l}] = 0.
\end{eqnarray}
Thus, the inclusion of the constituent quark operators completes the site-based
algebra of $U(2d) \times U(1)$ to $U(2d + 1)$.

Let us now consider a single rishon per link, i.e.\ ${\cal N}_{xy} = 1$. Then
$\mbox{det} U$ vanishes and the electric flux terms in the Hamiltonian of
Eq.(\ref{Hamiltonian}) (proportional to $g^2$ and ${g'}^2$ are trivial 
constants. Expressing the Hamiltonian in terms of glueball, meson, and 
constituent quark operators then leads to
\begin{eqnarray}
H&=&- t \sum_{\langle x y \rangle} 
\left(s_{xy} \psi_x^{i \dagger} U_{xy}^{ij} \psi_y^j + \mathrm{h.c.}\right) + 
m \sum_x s_x \psi_x^{i \dagger} \psi_x^i \nonumber \\
&-&\frac{1}{4 g^2}
\sum_{\langle w x y z \rangle} \left(U_{wx} U_{xy} U_{yz} U_{zw} + \mathrm{h.c.}\right)
\nonumber \\
&=&- t \sum_{\langle x y \rangle} 
\left(s_{xy} Q_{x,+k}^\dagger Q_{y,-k} + \mathrm{h.c.}\right) + 
m \sum_x s_x M_x \nonumber \\
&+&\frac{1}{4 g^2} \sum_{\langle w x y z \rangle} 
\Phi_{w,+k,-l} \Phi_{x,+l,+k} \Phi_{y,-k,+l} \Phi_{z,-l,-k},
\end{eqnarray}
where $y = x + \hat k$, $z = y + \hat l$, and $w = x + \hat l$. Remarkably, this
expresses the Hamiltonian entirely in terms of $U(N)$ gauge invariant 
operators. Still, it should be noted that the number of rishons per link is
conserved, i.e.\ $[H,{\cal N}_{xy}] = 0$. Also note that
\begin{equation}
{\cal N}_{xy} = \Phi_{x,+k,+k} + \Phi_{y,-k,-k}.
\end{equation}
This framework is well suited for exact diagonalization studies because it 
explicitly eliminates gauge variant states.

Under the various global symmetries the glueball, meson, and constituent quark
operators transform as
\begin{eqnarray}
&&^{S_k} \Phi_{x,\pm l,\pm m} = \Phi_{x+\hat k,\pm l,\pm m}, \nonumber \\
&& ^{S_k} M_x = M_{x+\hat k}, \ ^{S_k} Q_{x,\pm l} = Q_{x+\hat k,\pm l}, \nonumber \\
&&^{C_k} \Phi_{x,\pm l,\pm m} = 
- s_{\pm l} s_{\pm m} (\Phi_{x+\hat k,\pm m,\pm l} - N \delta_{\pm l,\pm m}), 
\nonumber \\
&&^{C_k} M_x = - M_{x+\hat k} + N, \nonumber \\
&&^{C_k} Q_{x,\pm l} = \mp (-1)^{x_1 + \dots + x_k} Q^\dagger_{x+\hat k,\pm l}, \nonumber \\
&&^P \Phi_{x,\pm l,\pm m} = \Phi_{-x,\mp l,\mp m}, \nonumber \\
&& ^P M_x = M_{-x}, \ ^P Q_{x,\pm l} = Q_{-x,\mp l}.
\end{eqnarray}
Here the sign $s_{\pm l} = \pm 1$ is the sign in $\pm l$. The above symmetry 
transformations follow directly from the transformation properties of the 
fermions and of the rishons (Eq.(\ref{rishonsym})). It is straightforward to 
show that they leave the commutation relations between the glueball, meson, and 
constituent quark operators invariant.

\section{Supplementary material II: Non-Abelian lattice gauge theories in alkaline-earth gases: 
the effect of imperfections in atomic implementations and STRUCTURE OF OPTICAL LATTICES }

\subsection{Introduction}

Imperfections in atomic physics implementations have to be carefully analyzed 
in order to understand the minimal experimental requirement to perform accurate
quantum simulations. Here, we discuss how typical experimental errors affect 
the precision of quantum simulations of non-Abelian lattice gauge theories in 
alkaline-earth setups, considering the paradigmatic case study of $U(2)$ 
quantum link models. In the next section, we briefly describe our model 
Hamiltonian, its experimentally relevant perturbations, and a set of 
observables quantitatively describing the departure from gauge invariance. In 
the last section, we present numerical results based on Lanczos diagonalization 
of small systems to underpin the validation issue at a quantitative level. 

\subsection{Microscopic model Hamiltonian, imperfections, and observables}

For simplicity, here we consider a $(1+1)D$ $U(2)$ QLM with two rishons per
link. As discussed in the main text, the QLM Hamiltonian $H$ is induced in 
second order perturbation theory from the microscopic Hamiltonian
\begin{eqnarray}\label{H_micro_I}
\widetilde{H} &= &
U \sum_x (n_{x,+} + n_{x+1,-} - 2)^2+ m \sum_x (-1)^x \psi^{i \dagger}_x \psi^i_x \nonumber \\
&+& \tilde t \sum_x (c^{i \dagger}_{x,+} \psi^i_x + c^{i \dagger}_{x,-} \psi^i_x +
\text{h.c.}) ,
\end{eqnarray}
where $n_{x,\pm} = c^{i \dagger}_{x,\pm} c^i_{x,\pm}$. This Hamiltonian is manifestly
$SU(2)$ gauge invariant, i.e.\ $[\tilde{H},G^a_x] = 0$. Unlike the QLM without the
$\text{det}U$ term, it is, however, not necessarily also $U(1)$ gauge invariant.
Also the constraint of ${\cal N}_{xy} = 2$ rishons per link will only be
satisfied approximately for sufficiently large $U$.

In an experimental realization with ultracold alkaline-earth atoms in an
optical lattice, one may expect some deviations from the ideal microscopic 
Hamiltonian. While interactions do not need any particular fine-tuning in our 
implementation, since the difference in scattering lengths for the different 
$m_I$ states are negligible compared to all other energy scales, 
different tunneling rates and local potentials require accurate lattice design, 
and may be subject to experimentally relevant imperfections. Such imperfections 
may give rise to color-dependent tunneling rates $\tilde t_i$. The actual 
Hamiltonian of a slightly imperfect experimental realization may take the form
\begin{eqnarray}
\widetilde{H}_{\text{imp}} &=&U \sum_x  (n_{x,+} + n_{x+1,-} - 2)^2 + m \sum_x (-1)^x \psi^{i \dagger}_x \psi^i_x  \nonumber \\
&+& \sum_x \tilde t_i (c^{i \dagger}_{x,+} \psi^i_x + c^{i \dagger}_{x,-} \psi^i_x +  \text{h.c.})  \nonumber \\
&+& 2 \Delta U \sum_x  (n_{x,+} - 1)(n_{x+1,-} - 1).
\end{eqnarray}
It should be noted that repeated indices are still summed. Due to the 
color-dependent hopping terms $\tilde t_i$, this Hamiltonian is no longer 
$SU(2)$ gauge invariant. The $\Delta U$ term, resulting from different overlaps 
of the Wannier functions in the partially overlapping rishon-sites, provides a 
source for the implementation of the $\text{det}U$ term in the $SU(2)$ model,
but should be viewed as an imperfection in the $U(2)$ context. In the 
following, we particularly focus on the effect of different tunneling rates for 
the two color species, investigating to what extent they can affect gauge 
invariance. Other sources of imperfections, such as trapping potentials
and small corrections to the interaction couplings, are expected to slightly shift
the effective parameters. Nevertheless, let us point out that, even in 
presence of explicitly gauge variant terms, the low-energy properties 
of the system are expected to be robust (see main text).

\subsection{Accuracy of gauge invariance: relevant observables}

In this section we investigate how deviations from a fixed rishon number per
link in the microscopic Hamiltonian $\widetilde H$ and color-dependent
tunneling rates in $\widetilde H_{\text{imp}}$ affect the accuracy of the gauge
symmetries.

\subsubsection{Abelian $U(1)$ gauge symmetry}

Let us first discuss the $U(1)$ symmetry generated by
\begin{equation}
G_x = n^1_x + n^2_x - \frac{1}{2}({\cal N}_{x,x+1} + {\cal N}_{x-1,x}),
\end{equation}
where
\begin{equation}
n^i_x = \psi^{i \dagger}_x \psi^i_x + c^{i \dagger}_{x,+} c^i_{x,+} +
c^{i \dagger}_{x,-} c^i_{x,-}
\end{equation}
counts the number of fermions of color $i$ on a basic building block
consisting of a quark- and the two adjacent rishon-sites forming a triple-well.
Since the fermion numbers $n^i_x$ are exactly conserved by construction, one 
should check whether the number ${\cal N}_{x,x+1} = 2$ of rishons per link 
remains constant. Deviations from this constraint are quantified by
\begin{equation}
\langle {\cal N}_{x,x+1} \rangle = 2+\Delta_{\text{link}},
\end{equation}
which should be small for sufficiently large $U$.

\subsubsection{Non-Abelian $SU(2)$ symmetry}

We now focus on the non-Abelian part of the symmetry group, $SU(2)$,
and check the expectation values of the generators
\begin{equation}
G^a_x = \psi^{i \dagger}_x \sigma^a_{ij} \psi^j_x + 
c^{i \dagger}_{x,+} \sigma^a_{ij} c^j_{x,+} +
c^{i \dagger}_{x,-} \sigma^a_{ij} c^j_{x,-},
\end{equation}
which should be zero in case of exact gauge symmetry. Here, $\sigma^a$ are the 
Pauli matrices. Notice that the diagonal $U(1)$ subgroup of the $SU(2)$ 
symmetry is also automatically fulfilled. This is because the generator
\begin{equation}
G^3_x = n^1_x - n^2_x 
\end{equation}
always has a vanishing expectation value by initial state preparation: the 
number of particles assigned to each basic building block (consisting of a 
quark- and two adjacent rishon-sites forming a triple-well) is 
strictly conserved. Here, we focus on the generator $G^1_x$ and calculate the
expectation value in the ground state of $H_{\text{imp}}$ (with $\Delta U = 0$)
\begin{equation}
D_1 = \frac{1}{L} \sum_x \langle (G_x^1)^2 \rangle.
\end{equation}

\subsection{Numerical results}

In this section, we present numerical results based on the previous theoretical
considerations.

\subsubsection{$U(1)$ rishon number symmetry: validation}

\begin{figure}[t]
\begin{center}
\includegraphics[width = 0.68\columnwidth]{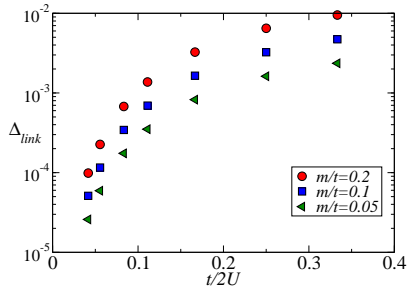}
\caption{Deviations from the conserved number of rishons as a function of 
$t/2U$ for different values of $m/t$.
\label{fig_U1}}  
\end{center}
\end{figure}

In Fig.~\ref{fig_U1}, we present exact diagonalization results for a system 
with 12 sites (4 quark- and 8 rishon-sites). This shows how the number of 
rishons per link, indicating the accuracy of the $U(1)$ symmetry, is very close 
to the desired value even for relatively small values of the interaction 
strength $U$. This is similar to the previously studied $U(1)$ case.

\begin{table}[hb]
\begin{center}

\begin{tabular}{|c|c|c|c|}
\hline
$D_1$ & $U = 3\tilde t$ & $U = 6 \tilde t$ & $U = 12 \tilde t$ \\
\hline
\hline
$z=0.5$&  $7.86 \cdot 10^{-3}$ & $9.34 \cdot 10^{-3}$ & $9.38 \cdot 10^{-3}$ \\ 
\hline
$z=0.4$&  $3.05 \cdot 10^{-3}$ & $3.35 \cdot 10^{-3}$ & $3.06 \cdot 10^{-3}$ \\ 
\hline
$z=0.3$&  $1.15 \cdot 10^{-3}$ & $1.15 \cdot 10^{-3}$ & $8.92 \cdot 10^{-4}$ \\ 
\hline
$z=0.2$&  $3.76 \cdot 10^{-4}$ & $3.46 \cdot 10^{-4}$ & $2.19 \cdot 10^{-4}$ \\ 
\hline
$z=0.1$&  $7.59 \cdot 10^{-5}$ & $6.71 \cdot 10^{-5}$ & $3.56 \cdot 10^{-5}$ \\ 
\hline
$z=0.01$& $6.7 \cdot 10^{-7}$  & $6.0 \cdot 10^{-7}$  &  $3.1 \cdot 10^{-7}$ \\ 
\hline
\end{tabular}
\caption{Discrepancy from exact gauge symmetry in the microscopic model as a 
function of $z$ for different inter-particle interactions $U$.}
\label{tab_error}
\end{center}
\end{table}

\subsubsection{$SU(2)$ symmetry: validation}

In Table \ref{tab_error}, we list typical values of $D_1$ as a function of the 
interaction strength $U$ (still considered $SU(2)$ symmetric) for approximate 
tunneling symmetries, where $\tilde t_2 = \tilde t$, 
$\tilde t_1 = (1-z)\tilde t$. Remarkably, even for relatively large values of 
$z \simeq 0.1$, errors due to approximate gauge invariance are extremely small, 
well beyond all other possible imperfections such as, e.g., the effect of the 
trapping potential or the space-dependent tuning of optical Feshbach resonances.

\subsection{Further remarks on the lattice structure}

Here we illustrate in a schematic way how the combination of 
triple-wells and the additional potential off-sets required
to impose the rishon constraint discussed in the previous subsection 
emerge from the lattice structure. For the sake of simplicity, 
we focus here on the 1D case, although the results generalize
straightforwardly in 2D and 3D. 

We define a set of color indices, $j=1, 2,...,N$, for the fermions.
We denote with $x$ the triple-well label, and with $x, x+, x-$ the site in the 
center of the well, on the right and on the left, respectively, according 
to the notation in Fig. 2a of the main text and the previous 
discussion.
The corresponding potential off-sets are then
\begin{eqnarray}
H_{\textrm{off}}&=&\sum_{j}\sum_{x~odd} \mu_{I} n^j_{x}+
\sum_{j}\sum_{x~even} \mu_{II} n^j_{x}+\nonumber\\
&+&\sum_{j}\sum_{x~odd}4\mu_{III}[n^j_{x,+}+n^j_{x,-}]+\nonumber\\
&+&\sum_{j}\sum_{x~even}4\mu_{III}[n^j_{x,+}+n^j_{x,-}].
\end{eqnarray}
The lattice has to realize a periodic structure of this kind,
either holographically or by employing a proper combination of laser
beams: in the 1D case, a combination of three lattice beams, just one of 
them being state-dependent, provides almost full tunability
of the microscopic parameters in $H_{\textrm{off}}$, although simpler,
less flexible choices might be employed for specific cases.
The corresponding microscopic terms in Eq.~\eqref{H_micro_I}
then read $m=(\mu_{I}-\mu_{II})/2, U=\mu_{III}$.
 Arrays of double-wells are routinely managed in current experiments, 
and the triple-well structure should constitute a variant of the same setups.

\end{document}